\documentstyle[12pt,aasms4,flushrt]{article}
\def\etal{{\it et al.~}}
\def\eg{{\it e.g.~}}

\def\kms{km~s$^{-1}$~}

\def\W50{W$_{50}$~}
\def\degree{$^{\circ}$~}
\def\lsim{~\rlap{$<$}{\lower 1.0ex\hbox{$\sim$}}}

\def\iras{{\it IRAS}~}
\def\oi{{\it Optical-IRAS}~}

\def\void#1{{}}
\begin{document}

\title {The Mass Distribution in the Nearby Universe}
\author{Luiz N. da Costa}
\affil {European Southern Observatory, Karl--Schwarzschild--Str. 2, D--85748
Garching b. M\"unchen, Germany}
\affil {Observat\'orio Nacional, Rua Gen. Jos\'e Cristino 77, 
S\~ao Cristov\~ao, Rio de Janeiro, Brazil}
\author {Wolfram Freudling}
\affil {Space Telescope--European
 Coordinating Facility and European Southern Observatory,
Karl--Schwarzschild--Str. 2, D--85748 Garching b. M\"unchen, Germany}
\author {Gary Wegner}
\affil {Dept. of Physics and Astronomy, Dartmouth College, Hanover, 
NH 03755}
\author {Riccardo Giovanelli,  Martha P. Haynes}
\affil{Center for Radiophysics and Space Research
and National Astronomy and Ionosphere Center\altaffilmark{1},
Cornell University, Ithaca, NY 14853}
\author {John J. Salzer}
\affil {Dept. of Astronomy, Wesleyan University, Middletown, CT 06457}
\altaffiltext{1}{The National Astronomy and Ionosphere Center is
operated by Cornell University under a cooperative agreement with the
National Science Foundation.}
\endtitlepage

\begin{abstract}  

We present a new reconstruction of the mass density and the peculiar
velocity fields in the nearby universe using recent measurements of
Tully-Fisher distances for a sample of late spirals.  We find
significant differences between our reconstructed fields and those
obtained in earlier work: overdensities tend to be more compact while
underdense regions, consisting of individual voids, are more
abundant. Our results suggest that voids observed in redshift surveys
of galaxies represent real voids in the underlying matter
distribution.  While we detect a bulk velocity of $\sim$ 300 \kms,
within a top-hat window 6000
\kms in radius, the flow is less coherent than previously claimed,
exhibiting a bifurcation towards the Perseus-Pisces and the Great
Attractor complexes.  This is the first time that this feature is seen
from peculiar velocity measurements.  The observed velocity field
resembles, more closely than any previous reconstruction, the velocity
field predicted from self-consistent reconstructions based on all-sky
redshift surveys. This better match is likely to affect estimates of the
parameter $\beta = \Omega^{0.6}/b$ and its uncertainty based on
velocity-velocity comparisons.

\end{abstract}

\section{Introduction}

Fluctuations in the mass distribution on intermediate scales can be
probed by peculiar velocity measurements, by measurements of
anisotropies in the cosmic background radiation at the 1\degree scale
and by distortions of galaxy images induced by weak field lensing.
Despite several pitfalls, the peculiar motion approach still remains
the most direct and the easiest to interpret, provided that
large--scale structures grow gravitationally and fluctuations in the
mass distribution induce the observed peculiar velocities of galaxies.
Mapping the peculiar velocity field can thus provide information
useful to constrain the mass power--spectrum, the relation between the
galaxy and matter distributions, and the value of the cosmological
density parameter $\Omega$ on large scales.

In recent years, considerable effort has been made to understand the
nature of the peculiar velocity field and the determination of the
parameter $\beta = \Omega^{0.6} /b $, where $b$ is the linear biasing
parameter. Measurements of $\beta$ have been attempted either from the
comparison of the observed fluctuations in the galaxy distribution
with those obtained from dynamical studies or from the direct
comparison of the measured peculiar velocity field to that predicted
from dynamically self-consistent reconstruction of the density field
from all--sky redshift surveys (for a review see Dekel 1994). These
estimates based on peculiar velocity measurements have generally led
to higher values than those obtained on small scales by other methods
(\eg Strauss \& Willick 1995). The reason for this inconsistency
remains an open question.

Until recently, most of the work was based on compilations of distance
measurements by different authors which were sparse, inhomogeneous and
probed a shallow volume in a non-uniform way (Bertschinger \etal 1990,
Dekel 1994). Although several new peculiar velocity surveys have been
carried out (Willick 1990, Courteau \etal 1993, Mathewson, Ford \&
Buchhorn 1992b, Wegner \etal 1996), none covers the whole sky
uniformly.  Efforts to homogenize the existing datasets are currently
underway (Willick \etal 1996) but error estimates for the distances of
individual galaxies are likely to continue to be uncertain. This
directly affects the estimation of the bias which might, in the case
of the inhomogeneous Malmquist bias correction, lead to spurious
detection of density fluctuations.

The need for a sample with uniform sky coverage and homogeneous
measurements has prompted us to carry out an independent
redshift-distance survey of Sbc-Sc galaxies, based on I-band
Tully-Fisher (TF) distances, covering essentially the whole sky (dec
$\ge$ -45\degree and $|b| \ge$ 10\degree).  The sample has
well-defined selection criteria and new measurements were carried out
in order to assure the homogeneity of the data. This data set was
complemented with galaxies drawn from Mathewson, Ford \& Buchhorn
(1992b, MFB), appropriately converted to our magnitude and width
scale, yielding a combined sample of about 1300 field galaxies
(hereafter SFI sample) and 500 galaxies in clusters.  Details on sample
selection and the procedure adopted to combine the two datasets can be
found in Giovanelli \etal (1994, 1995a,b). Here, we present a summary
of the results of a reconstruction of the three-dimensional velocity
and density fields obtained from the SFI sample. We use the 3D
reconstruction method primarily to illustrate the nature of the sample
as compared to those currently available. A more quantitative analysis
will be the subject of future papers.  In section 2, we briefly
describe the reconstruction method and discuss the approach used to
correct for the various biases that affect the distance estimates. Our
main results are presented in section 3, while in section 4 we
summarize our conclusions.

\section{Analysis of the SFI Sample}

Distances of galaxies were estimated using the direct I-band TF
relation derived from the combination of data for 24 clusters
(Giovanelli \etal 1996 a,b). However, errors in the distance
measurements lead to biases which directly affect the peculiar
velocity field and thus the derived density field.  The biases result
from the coupling of uncertainties in the distance estimate with
inhomogeneities in the galaxy distribution, and the selection of the
sample. The latter is particularly important for the Sc sample because
the adopted redshift-dependent selection criteria (\eg Giovanelli
\etal 1994) implies that near each redshift limit outflowing galaxies
are preferentially excluded from the sample. This effect becomes
dominant at large distances leading to a spurious systematic infall of
galaxies at the outer edge of the surveyed volume. In order to deal
with this problem we have resorted to a numerical approach to estimate
the bias field at each point in estimated distance space (Freudling
\etal 1995). The method, which is based on mock catalogs with
characteristics similar to the data, allows for the simultaneous
correction of the homogeneous and inhomogeneous Malmquist bias, and
biases introduced by selection effects. The mock catalogs are
generated from the distribution of optically-selected galaxies in real
space. This distribution was obtained using an iterative method of
reconstructing a dynamically self-consistent distribution of galaxies
from all-sky redshift surveys (\oi sample) using a quasi-linear theory
(Freudling, da Costa \& Pellegrini 1994).

The bias correction is sensitive to the value adopted for the scatter
in distance estimates. Attempts to use the scatter as determined for
the TF relation of galaxies in clusters led to an overcorrection of
the velocity field. In order to overcome this problem we have also
used the mock catalogs to determine the scatter in the distance
estimates of field galaxies, taking advantage of the homogeneity of
the sample. This was done by comparing the observed radial peculiar
velocities at large distances with those of mock catalogs, since these
velocities are primarily produced by biases which depend on the
assumed scatter. The scatter of the mock samples was adjusted so that
the radial variation of the peculiar velocities at large distances in
the mock catalog match the observations.  The derived scatter of
$\lsim 0.4$ mag depends on the HI line-width, and  is in any case
smaller than that obtained for galaxies in clusters (Giovanelli \etal
1996a). A more extensive discussion of the motivation and the details
for the adopted procedure will be presented in da Costa \etal (1996).

Distances were calculated from the TF relation and the radial
component of the peculiar velocity $u = cz_{CMB} -d_c$, where $d_c$ is
the ``fully'' corrected distance estimate.  Galaxies in clusters were
assigned to a single object. A small fraction of galaxies ($\sim$ 6\%)
with small line widths ($\log w <2.25$) were discarded because of the
large fractional errors in the measurement of their widths. The final
input sample used in the reconstruction of the density field consists
of 1234 independent points.

The computed distances and peculiar velocities are the input data to
reconstruct the density and three-dimensional peculiar velocity
fields.  For the reconstruction, we use the standard assumption of
irrotational flow which allows calculation of the scalar velocity
potential field from the integral along radial paths of the
line-of-sight component of the peculiar velocity.  Our algorithm is
similar to that discussed by Dekel, Bertschinger \& Faber (1990). We
adopt a grid in spherical coordinates of 20 equal shells out to a
maximum radius of 7500 \kms and 64 latitude and longitude circles. The
measured radial peculiar velocities were smoothed using a Gaussian
tensor window with a fixed smoothing scale of $R_w$ = 900 \kms.  Our
methodology was extensively tested using mock catalogs with the same
number of objects and selection criteria as the data, which were also
used to determine the uncertainties in the reconstruction (da Costa
\etal 1996).

\section{Results}
 
The reconstructed density fluctuation field $\delta$ along the
Supergalactic plane on a cubic grid of spacing 500 \kms for a fully
corrected model is displayed in figure 1, where we show in the top
plane density contours in $\delta= 0.2$ intervals, in the middle plane
surface density maps with the contour level proportional to the
density contrast $\delta$ and in the bottom plane density contrast
maps. We use a Cartesian supergalactic coordinate system with SGX-axis
pointing to $L=0$\degree and the SGY-axis pointing nearly to the North
Galactic Pole.

>From a visual inspection of the map one can recognize some well-known
structures in the nearby volume such as the Great Attractor (GA,
SGX~$\sim$~-2000~\kms, SGY~$\sim$~-500~\kms), the Perseus-Pisces (PP)
complex (SGX~$\sim$~6000~\kms, SGY~$\sim$~-1000~\kms) and the large
void (SGX~$\sim$~2500~\kms, SGY~$\sim$~0~\kms) between the GA and the
PP.  We can also see traces of the Coma/A1367 supercluster along
SGY~$\sim$~7000~\kms, and the Cetus region (SGX~$\sim$~500~\kms,
SGY~$\sim$~-6000~\kms). These structures are located roughly at the
same position as in  the \iras reconstructed density field (\eg Strauss
\& Willick 1995)

In comparison to earlier work (\eg Dekel 1994), better sampling of the
northern hemisphere apparently leads to significant differences in the density
field especially in the region not covered by MFB.  Some important
differences can be seen: 1) in the foreground of the Coma/A1367
supercluster, where now a large void is detected; 2) in the PP region,
which appears as a compact high density peak; and 3) in the Great
Attractor region, which looks less prominent. Examination of the
reconstructed density field in three-dimensions also suggests that the
GA consists of different sub-structures, much in the way seen in
redshift surveys (\eg Willmer \etal 1995).

A striking feature of our reconstructed density field is the existence
of several voids surrounded by tenuous structures.  Among them is the
nearby void separated from the voids in the foreground of the
Coma/A1367 supercluster by a small amplitude coherent structure, which
we identify as the Local Supercluster. It is, as expected, much less
prominent than other structures within the volume. Other prominent
voids are the void behind the GA (SGX~$\sim$~-6500~\kms,
SGY~$\sim$~1500~\kms) and the Sculptor void (SGX~$\sim$~-2000~\kms,
SGY~$\sim$~-6000~\kms). Both voids have been detected in galaxy
redshift maps (\eg da Costa \etal 1988, Geller \& Huchra 1989) but are
now seen in the mass distribution with density contrasts reaching
$\delta \lsim -0.6$, as indicated by the contours in the top plane.

Several other positive and negative fluctuations are seen near the
outer edge of our survey volume. Comparison with the \oi and {\it
IRAS} density fields indicate that most are likely to be real
structures. However, it is important to emphasize that the error in
the density field derived from mock catalogs, which is smaller than
0.2 within 5000 \kms, grows rapidly with radius. At 6000 \kms the
error is estimated to be about 0.4, corresponding to two contour
levels. Therefore, the amplitude and location of structures beyond
5000 \kms are very poorly determined. This is also true for the
Coma/A1367 and PP superclusters. However, although PP itself is poorly
mapped, its influence in the velocity field extends over a large
volume and into the region where the errors are small and the results
insensitive to the details of the inhomogeneous bias correction.

The density field reconstructed with the SFI sample is characterized by
overdensities which are compact and underdensities which are large, have
regular shape and have a high density contrast.  This is in marked
contrast to earlier reconstructions. Our results suggest the existence
of real voids in the matter distribution, qualitatively similar to those
observed in galaxy redshift surveys (\eg da Costa \etal 1994).  This
supports the contention that galaxies delineate real voids in the
underlying matter distribution which may impose severe constraints in
the amplitude of the mass power-spectrum (\eg Piran \etal 1993).

In Figure 2, we show the velocity field along the Supergalactic plane
superimposed to the density field shown in figure 1.  This figure
should be compared to a similar map presented by Dekel (1994).  In
contrast to Dekel's result our velocity field is qualitatively similar
to the predicted \iras velocity field (\eg Yahil 1988) showing a
region where the flow bifurcates towards the GA and towards the PP
complex. If true, this resolves a longstanding discrepancy between
predicted and measured velocity fields. Moreover, an infall into PP is
clearly observed while none is seen in Dekel's (1994)
reconstruction. This contrasts with the interpretation of Willick
(1990) who claims that `the principle feature of the velocity field in
the PP region is a coherent streaming towards the Local Group'.
Willick's smaller sky coverage of the PP region could well be the
cause of this difference.

On the opposite side of the figure, we find that the GA is less
important than originally believed. Instead of being responsible for
most of the motion in the nearby volume, there is strong evidence that it
is smaller and less massive than originally estimated by Lynden-Bell
\etal (1988). This is probably due to the larger surveyed volume and
better sampling of the PP region. Note, in particular, that the
amplitude of the backflow into the GA is small.  Instead, the flow
field in the upper left hand side of Figure 2 could be explained by a
mass concentration beyond the survey volume pulling the GA.  This
would explain the lack of an unambiguous signature for backside infall
in the survey of Burstein, Faber \& Dressler (1990). Finally, the
volume-weighted bulk velocity as measured within a top-hat window 6000
\kms in radius is about 300 \kms in the direction $L \sim$ 160 \degree and $B
\sim $ -20\degree , where $L$ and $B$ are the supergalactic coordinates.
These values are very close to those reported by Dekel (1994).
However, the flow shows an overall shear unlike the coherence
suggested by Mathewson, Ford and Buchhorn (1992a) and Courteau \etal
(1993).

There are several possible explanations for the observed differences
between the results of our reconstruction and those of earlier work
(\eg Dekel 1994). The uniform sampling over the sky and depth is
certainly a major factor, as pointed out above. Differences in the
assumed scatter properties and methods for inhomogeneous Malmquist bias
correction could also contribute to the position-dependent differences
observed.  Differences in the relative zero-points between different
data sets in the sample used by Dekel (1994) could be another reason
for the observed discrepancies.  A detailed comparison of the two
datasets will be required to better understand the reason for the
discrepancies.

\section{Conclusions}

The picture that emerges from our reconstruction brings together the
somewhat fragmentary view previously held. The main results may be
summarized as follows:

1- The positive fluctuations of the matter density field tend to be
more compact while well-defined voids are more abundant than those
obtained in previous reconstructions.

2- Comparison with redshift maps suggests that galaxies delineate real
voids in the mass distribution. These voids tend to be separated by
relative low-density structures which correspond to the filamentary
and wall--like structures observed in the galaxy distribution.

3- The GA is significantly less prominent than previously claimed.
There is also some indication for sub--structures.

4- An outside mass concentration may be pulling the GA. This induced
motion overshadows the backflow and explains the lack of convincing
evidence of such a flow.

5- The PP supercluster makes a significant contribution to the flow in
its neighborhood leading to an infall into the supercluster seen for
the first time from peculiar velocity measurements.

6- The measured mean bulk velocity within a top-hat window of
6000~\kms is about 300~\kms. The flow shows an overall shear
which may be consistent with the existence of a mass concentration
outside the observed volume.

Our results should affect conclusions drawn earlier regarding the
value of $\beta$, or at least the accuracy with which it can be
determined, because of the better agreement between the velocity fields
reconstructed from peculiar velocity measurements and from the galaxy
distribution. The results may also affect the conclusions regarding
the biasing of galaxies relative to the mass and the mass
power-spectrum. These topics will be pursued in more detail in future
papers.

\acknowledgements
We would like to thank A. Dekel for many useful discussions and to the
anonymous referee for his constructive comments.  LNdC
would like to thank the hospitality of the Institute d'Astrophysique
and the Hebrew University where part of this work was carried out. GW
thanks for the hospitality of the European Southern Observatory, NSF
Grant AST93-47714 and the Alexander von Humboldt foundation for
partial support.  We acknowledge the financial support from
U.S. National Science Foundation grants AST94--20505 to RG,
AST92--18038 and AST90--23450 to MPH and from the Research Corporation
to JJS.  The results presented in this paper are based on observations
carried out at the European Southern Observatory (ESO), the National
Astronomy and Ionosphere Center (NAIC), the National Radio Astronomy
Observatory (NRAO), the Kitt Peak National Observatory (KPNO), the
Cerro Tololo Interamerican Observatory (CTIO), the Observatory of
Paris at Nan\c cay and the Michigan--Dartmouth--MIT Observatory
(MDM). NAIC, NRAO, KPNO and CTIO are respectively operated by Cornell
University, Associated Universities, inc., and Associated Universities
for Research in Astronomy, all under cooperative agreements with the
National Science Foundation.
\vfil\eject

\vfil \eject

\noindent{\bf Figure Captions}

\figcaption[] {(Plate) - Density field along the
Supergalactic plane reconstructed from the SFI sample. We show density
contours in intervals of $\delta = 0.2$, surface density maps with the
height proportional to $\delta$ and density constrast maps. Bright
yellow represents overdensities and dark blue underdense regions,
respectively.  See text or figure 2 for the identification of the
structures.}

\figcaption[] {Projected components of the three-dimensional peculiar
velocity field along the Supergalactic plane superimposed to the
density field. The contour spacing is 0.2 in $\delta$, with the
heavy solid line indicating $\delta=0$. The coordinate system is as
described in the text. The major structures seen in the map are the
Great Attractor, the Perseus-Pisces complex, the Cetus region and the
foreground of the Coma/A1367 supercluster.}


\begin{references}
\reference{} Bertschinger, E., Dekel, A., Faber, S.
M., Dressler. A. \& Burstein, D. 1990, \apj~ 364, 370.
\reference{} Courteau, S., Faber, S.M., Dressler, A. and Willick, J.A.
1993, \apj ~412 L51.
\reference{} Burstein, D., Faber, S. M. \& Dressler, A. \apj ~354, 18
\reference{} da Costa, L.N., Pellegrini, P.S., Sargent,W.L.W., Tonry,
J., Davis, M., Meiksin, A., Latham, L., Menzies, J. \& Coulson, I.,
1988, ApJ 327, 544 
\reference{}da Costa, L.N., Geller, M.J., Pellegrini, P.S., Latham, D.W.,
Fairall, A.P., Marzke, R.O., Willmer, C.N.A., Huchra, J.P., Calderon, J.H.,
Ramella, M., \& Kurtz, J. 1994, ApJ 424, L1
\reference{} da Costa \etal 1996, in preparation
\reference{} Dekel, A.,  Bertschinger, E. \& Faber, S. M.
1990, \apj ~364, 349 
\reference{} Dekel, A. 1994, \araa\,  32, 371
in the Universe, ed. V. Rubin \& G. V. Coyne (Princeton: Princeton
Univ. Press), 116
\reference{} Freudling, W., da Costa, L. N. \& Pellegrini, P. S. 1994,
 \mnras ~268, 943 
\reference{} Freudling, W., da Costa, L. N., Wegner, G.,
Giovanelli, R, Haynes, M. P. \& Salzer, J.J. 1995, \aj  ~110, 920  
\reference{} Geller, M.G. \& Huchra, J.P. 1989, Science 246, 897
\reference{} Giovanelli, R., Haynes, M.P., Salzer, J.J., Wegner, G.,
da Costa, L.N. \& Freudling, W. 1994, \aj ~107, 2036
\reference{} Giovanelli, R., Haynes, M.P., Salzer, J.J., Wegner, G.,
da Costa, L.N. \& Freudling, W. 1995a, \aj ~110, 1059
\reference {} Giovanelli, R., Haynes, M.P., Chamaraux, P., da Costa,
L.N., Freudling, W., Salzer, J.J. \& Wegner, G. 1995b, in {\it Examining the
Big Bang and Diffuse Background Radiations}, proc. of IAU Symp. nr. 168,
ed. by M. Kafatos, p. 183
\reference{} Giovanelli \etal 1996a, in preparation
\reference{} Giovanelli \etal 1996b, in preparation
\reference{} Lynden-Bell, D. Faber, S.M., Burstein, D., Davies, R.L., Dressler, A. Terlevick, R.J. \& Wegner, G.
1988, \apj ~326, 19
\reference{} Mathewson, D.S., Ford, V.L. \& Buchhorn, M. 1992a \apj ~389, L5
\reference{} Mathewson, D.S., Ford, V.L. \& Buchhorn, M. 1992b \apjs~81, 413
\reference{} Piran, T., Lecar, M., Goldwirth, D., da Costa, L. \& Blumenthal,
G. 1993, MNRAS ~265, 681
\reference{} Strauss, M. \& Willick, J. A. 1995, Physics Report 261, 271
\reference{} Wegner, G., Colless, M., Baggley, G., Davies, R.L., Bertschinger,
E., Burstein, D., McMahan Jr., R.K. \& Saglia, R.P. 1996, preprint.
\reference{} Willick, J. A. 1990, \apj ~351, L5
\reference{} Willick, J., Courteau, S., Faber, S. M., Burstein, D.,
Dekel, A. \& Kollat, T. 1996, ApJ, 457, 460 
\reference{} Willmer, C.N.A., da Costa, L.N., Pellegrini, P.S., Fairall, A.
\& Latham, D. 1995, \aj ~109, 61
\reference{} Yahil, A. 1988, in Large-Scale Sructuture of the
Universe,ed. V.C. Rubin G.V. Coyne (Princeton: Princeton University
Press) p. 219
\end{references}
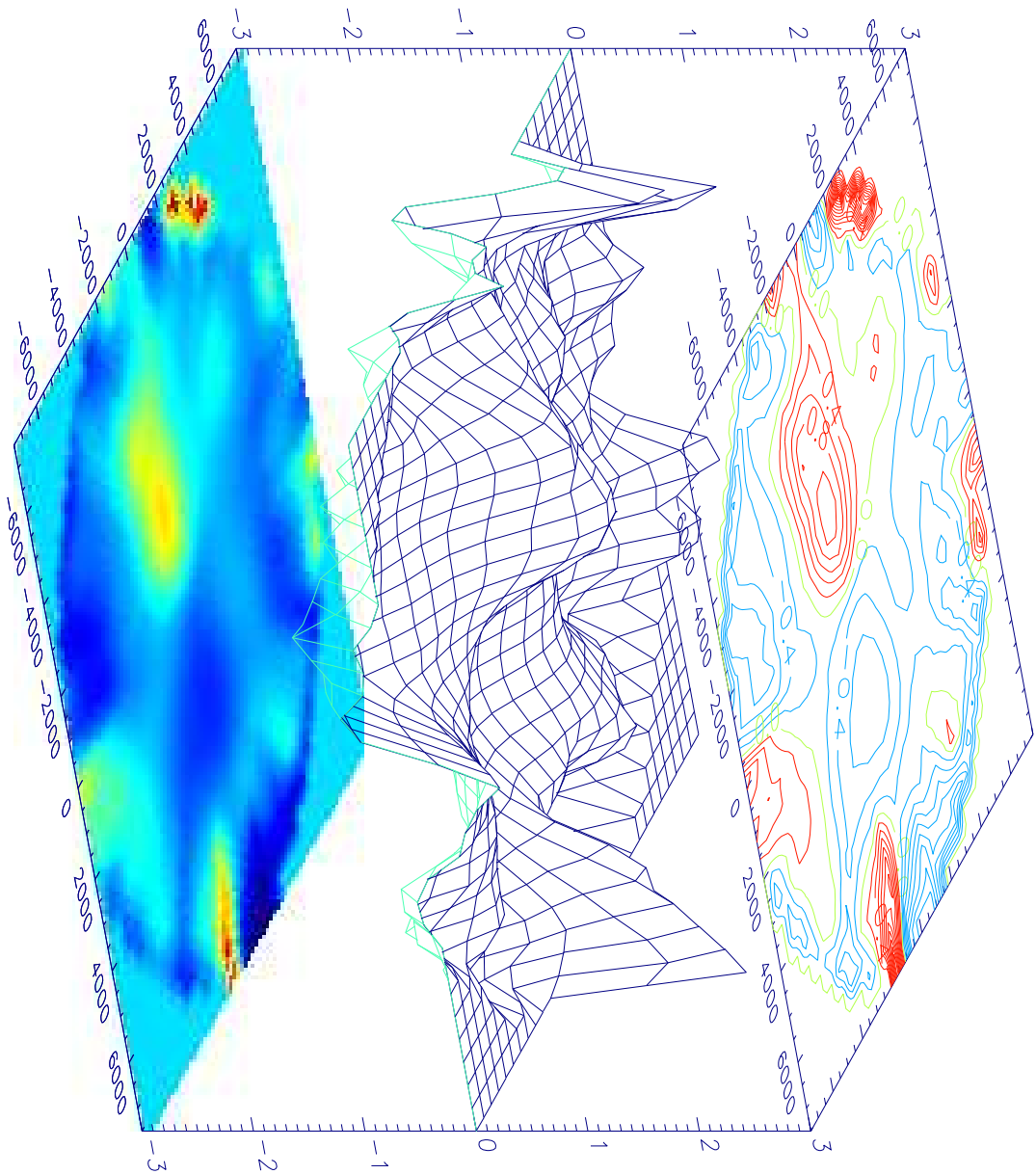
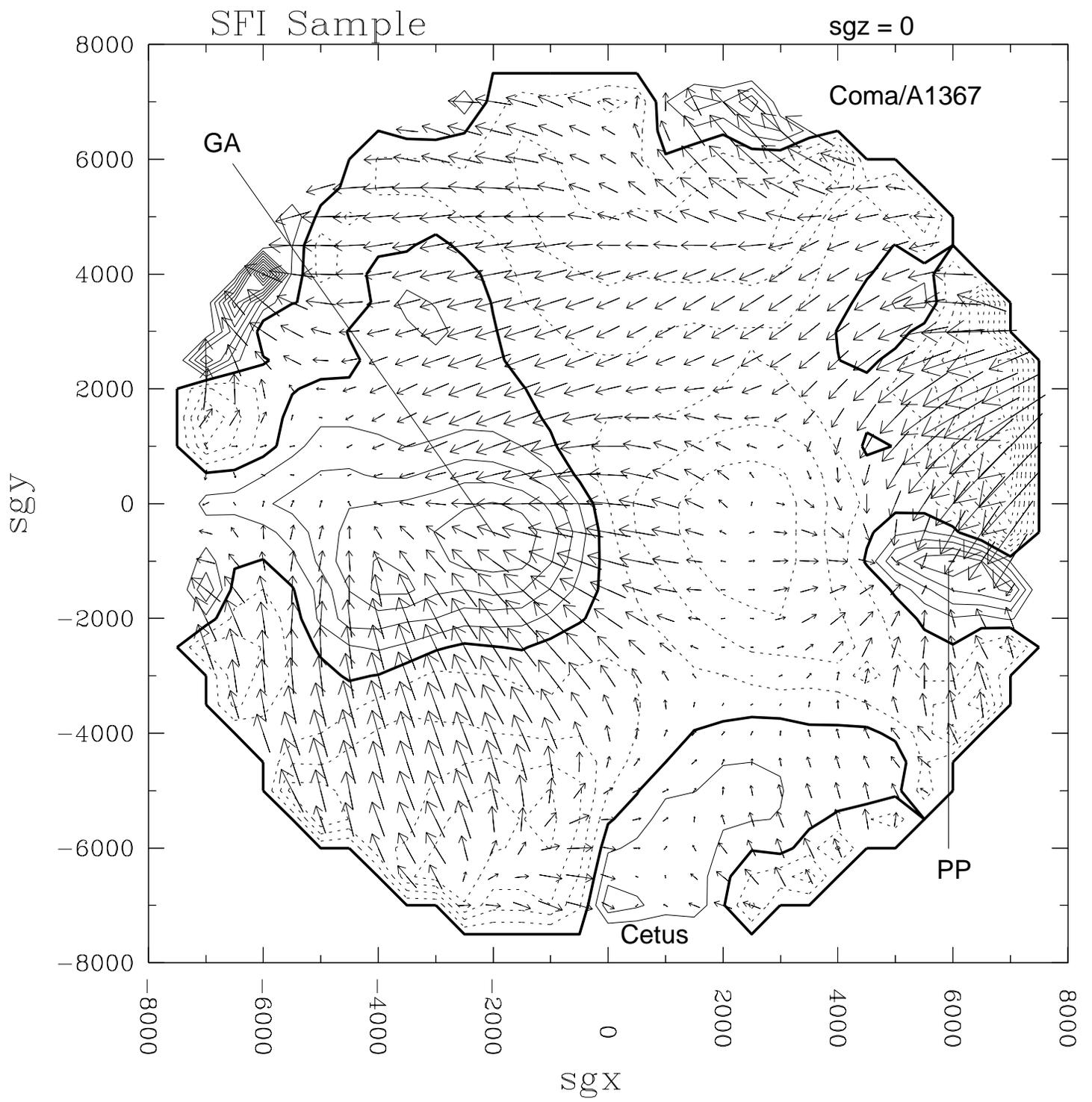
\end{document}